\documentclass[a4paper,11pt]{JHEP3}
\usepackage{amsmath}
\usepackage{epsf, amssymb}
\usepackage{epsfig}

\newcommand{\Real}{{\rm Re}}
\newcommand{\Imagi}{{\rm Im}}
\newcommand{\half}{\frac{1}{2}}
\newcommand{\quart}{\frac{1}{4}}
\newcommand{\ie}{{\it i.e.\ }}

\newcommand{\dT}{\int_{\cal F}\!\frac{d^2\tau}{(\Imagi\,\tau)^5}}
\newcommand{\dV}{\int_{\cal T}\prod_{r=1}^4 d^2v_r}
\newcommand{\dVdash}{\int_{\cal T}\prod_{r'=1}^4 d^2v_{r'}}
\newcommand{\chiprod}{\prod_{r<s}(\chi_{rs})^{\frac{1}{2}k_r \cdot k_s}}
\newcommand{\chiproddash}{\prod_{r'<s'}(\chi_{r's'})^{\frac{1}{2}k_{r'} \cdot k_{s'}}}
\newcommand{\ImTau}{\Imagi\,\tau}
\newcommand{\Imv}{\Imagi\,v}
\newcommand{\tr}{\mbox{Tr}}

\newcommand{\Rf}{{\mathcal{R}^4}}
\newcommand{\NSNS}{{$NS\otimes NS$}}

\newcommand{\rem}[1]{}

\title{The One-Loop $H^2R^3$ and $H^2(\nabla H)^2R$ Terms in the Effective Action}
\author{David M. Richards \\
        Department of Applied Mathematics and Theoretical Physics \\
        University of Cambridge \\
        Wilberforce Road, Cambridge CB3 0WA, United Kingdom.\\
        E-mail: \email{D.M.Richards@damtp.cam.ac.uk}}
\abstract{
We consider the one-loop $B^2h^3$ and $B^4h$ amplitudes in type II string
theory, where $B$ is the \NSNS\ two-form and $h$ the graviton, and
expand to lowest order in $\alpha'$. After subtracting diagrams due to
quartic terms in the effective action, we determine the presence and
structure of both an $H^2R^3$ and $H^2(\nabla H)^2R$ term. We show
that both terms are multiplied by the usual
$(t_8t_8\pm\frac{1}{8}\epsilon_{10}\epsilon_{10})$ factor.
}
\preprint{DAMTP-2008-61}

\begin{document}

\section{Introduction}
In a recent paper \cite{Richards:2008jg} we considered the one-loop
five-graviton amplitude, $h^5$, in type II string theory. From its
low-energy expansion we concluded that $R^5$ and $\nabla^2R^5$ terms
are absent from the effective action, but that $\nabla^4R^5$ terms are
present, where $R$ is the Riemann tensor. This paper considers similar
amplitudes but with \NSNS\ two-form potentials, $B$, replacing some of
the gravitons. In particular we consider the one-loop $B^2h^3$ and
$B^4h$ amplitudes and expand to lowest order in $\alpha'$. This
reveals the presence and tensor structure of $H^2R^3$ and $H^2(\nabla
H)^2R$ terms in the effective action, where $H$ is the field strength
associated with $B$.

For our purposes, the low-energy effective action is a functional of
the massless spectrum of string theory such that, even for string
loops, only {\it tree} diagrams are required to reproduce string
amplitudes. This is the sense in which the famous $R^4$ correction to
supergravity, as in for example \cite{Green:1997tv}, should be
interpreted. After expanding an amplitude for small $\alpha'$, new
terms in the effective action can only be determined after diagrams
due to previously-known terms are subtracted. For example, before we
can find $H^2R^3$ from the lowest-order expansion of the $B^2h^3$
amplitude, it is necessary to subtract diagrams involving quartic
effective action terms, such as the $(\nabla H)^2R^2$ term.

We calculate amplitudes using the light-cone gauge GS formalism, which
requires that $k^+$ vanishes for all external states. As a
consequence, there are certain terms, both in amplitudes and in the
effective action, that cannot be discovered. For example,
$\epsilon_{10}\epsilon_{10}$ terms with fewer than two contractions
between the epsilons, such as (3.13) in \cite{Peeters:2001ub}, will be
missed. Similarly, the $Bh^4$ amplitude, and hence the one-loop
$B\wedge t_8R^4$ term found in \cite{Vafa:1995fj}, will not be
found. However, all other terms, and in particular
$\epsilon_{10}\epsilon_{10}$ terms with at least two contractions
between the epsilons, will be seen.

The plan of this paper is as follows. In section \ref{sec:4ptAmps} we
review the one-loop four-graviton amplitude, the associated $R^4$ term
in the effective action, and its extension to include \NSNS\ two-forms
found in \cite{Gross:1986mw}. The $B^2h^3$ amplitude is calculated in
section \ref{sec:B2h3amp} and then expanded to lowest-order in
$\alpha'$. Section \ref{sec:EA} is concerned with expanding the
quartic effective action and calculating the relevant diagrams. After
these are subtracted, the remaining terms are covariantised to
discover a new $t_8t_8H^2R^3$ term. The whole analysis is extended to
the $B^4h$ case in section \ref{sec:B4h}, which results in a new
$t_8t_8H^2(\nabla H)^2R$ term. Finally, section \ref{sec:eeterms}
pays closer attention to $\epsilon_8$ terms in the amplitudes. This
shows that the $t_8t_8$ in both $H^2R^3$ and $H^2(\nabla H)^2R$ should
be generalised to $(t_8t_8\pm\frac{1}{8}\epsilon_{10}\epsilon_{10})$,
with $+/-$ for IIB/IIA, where one pair of indices is contracted
between the epsilon tensors. Throughout we will use a metric with
signature $\{-,+,+,+,\ldots\}$ and will often set $2\alpha'=1$.

\section{The Effective Action from Four-Point Amplitudes} \label{sec:4ptAmps}
Before considering amplitudes for five states, we review terms in the
effective action which arise from four-particle amplitudes involving
gravitons and \NSNS\ two-forms. For the case of four gravitons the
one-loop amplitude is well-known to be given by \cite{Green:1987mn}
\begin{equation} \label{eq:4gravAmp}
   A_{4h} = \hat{K} \dT \int_{\cal T}\prod_{r=1}^3 d^2v_r \chiprod,
\end{equation}
with
\begin{equation}
  \hat{K} = t_8^{a_1b_1\cdots a_4b_4} t_8^{c_1d_1\cdots c_4d_4}
  k^1_{a_1}k^1_{c_1}h^1_{b_1d_1} k^2_{a_2}k^2_{c_2}h^2_{b_2d_2}
  k^3_{a_3}k^3_{c_3}h^3_{b_3d_3} k^4_{a_4}k^4_{c_4}h^4_{b_4d_4},
\end{equation}
where the four gravitons have polarisations $h^r_{a_rb_r}$ and
momenta $k^r_{a_r}$, and $r$ ranges from $1$ to $4$. Here $v_r$ are
the positions of the vertex operators on the torus and their integrals
are taken over the rectangular region $-\half<\Real v\leq\half,
-\half\ImTau<\Imv\leq\half\ImTau$, denoted by ${\cal T}$; whereas the variable $\tau$
parameterizes the modulus of the torus and so is integrated over a
fundamental domain of $SL(2,\mathbb{Z})$, denoted by ${\cal F}$. The function
$\chi_{rs}\equiv\chi(v_r-v_s,\tau)$ is a non-singular, doubly periodic
function of $v$ and $\bar{v}$ which is given explicitly in
\cite{Green:1987mn}. The $t_8$ tensor originates from a trace over
eight fermionic zero modes and can be written as a sum of an
$\epsilon_8$ tensor and sixty $\delta\delta\delta\delta$ tensors
\cite{Green:1987mn}. Often $t_8$ is defined without the $\epsilon_8$
tensor, especially when written in effective actions, and we will
clarify this issue later. However, for the four-graviton amplitude
this difference is not important since the $\epsilon$ parts vanish by
momentum conservation.

As shown in \cite{D'Hoker:1993ge, D'Hoker:1993mr, D'Hoker:1994yr},
the integrals in \eqref{eq:4gravAmp} only converge for $s=t=u=0$ where
$s$, $t$, $u$ are the usual four-particle Mandelstam variables defined
in \cite{Richards:2008jg}. Even for complex values of the momenta, the
convergence is only for purely imaginary values of $s$, $t$ and
$u$. The resolution is to analytically continue from the imaginary
axis to the entire complex plane. Only then can the amplitude be shown
to contain the correct massive poles and threshold cuts demanded by
unitarity.

The low-energy expansion of one-loop amplitudes can be quite involved
\cite{Green:1999pv, Richards:2008jg}, but since we only require the
expansion at lowest-order in $\alpha'$ the situation is much
simpler. To find the lowest-order expansion of \eqref{eq:4gravAmp} we
set $k_r\cdot k_s$ to zero for all $r$, $s$ giving
\begin{equation} \label{eq:4gravexp}
  A_{4h}|_{\alpha'^3} = \hat{K} \int_{\cal F}\!\frac{d^2\tau}{(\ImTau)^2} =
  \frac{\pi}{3}\hat{K},
\end{equation}
where the power of $\alpha'$ is, as throughout this paper, relative to
the Einstein-Hilbert term. It is trivial to covariantise this and find
the famous $t_8t_8R^4$ term in the effective action. If this one-loop
term is combined with the equivalent tree-level result found in
\cite{Gross:1986iv} then the $\alpha'^3$ term is given in Einstein
frame by
\begin{equation} \label{eq:R4}
  \alpha'^3\int d^{10}x \sqrt{-g}\,\left(2\zeta(3)e^{-3\phi/2} +
  \frac{2\pi^2}{3}e^{\phi/2}\right) \Rf,
\end{equation}
where $\Rf$ is shorthand for
\begin{equation} \label{eq:R4def}
  t_8^{a_1b_1a_2b_2a_3b_3a_4b_4}t_8^{c_1d_1c_2d_2c_3d_3c_4d_4}
  R_{a_1b_1c_1d_1}R_{a_2b_2c_2d_2}R_{a_3b_3c_3d_3}R_{a_4b_4c_4d_4},
\end{equation}
and where we have fixed the normalisation for the string S-matrix so
that the one-loop term contains an extra factor of $2\pi$ relative to
the tree-level term \cite{Green:2008uj}.

In the case of IIB it is possible to extend the $\Rf$ term to all
orders in the string coupling, even including non-perturbative
effects. It was shown in \cite{Green:1997tv, Green:1997di,
Green:1998by, Green:1997as} that the complete $\Rf$ action is given by
$\alpha'^3\int d^{10}x \sqrt{-g}\,Z_{3/2}(\tau,\bar\tau) \Rf$, where
$Z_{3/2}$ is a non-holomorphic Eisenstein series given by
\begin{align}
  Z_{3/2}(\tau,\bar\tau) &= \sum_{(m,n)\neq(0,0)}
  \frac{(\ImTau)^{3/2}}{|m\tau+n|^3} \notag\\
  &= 2\zeta(3)e^{-3\phi/2} +
  \frac{2\pi^2}{3}e^{\phi/2} \notag\\
  &\qquad + 4\pi\sum_{k\neq
  0}\mu(k)e^{-2\pi(|k|e^{-\phi}-ikC^{(0)})}k^{1/2}
  (1+\frac{3}{16\pi|k|}e^{\phi}+\ldots),
\end{align}
with $\mu(k)=\sum_{d|k}d^{-2}$. Here $\tau$, which should not be
confused with the modular parameter in one-loop amplitudes, is the
usual combination of the Ramond-Ramond scalar and the dilaton, $\tau =
C^{(0)} + ie^{-\phi}$. The expansion shows that there are no
perturbative contributions beyond one-loop, but that there are an
infinite sum of single D-instanton terms, with characteristic
$e^{1/g}$ behaviour, which were first studied in \cite{Green:1997tv}.

Four-particle amplitudes containing \NSNS\ two-forms were first studied
in \cite{Gross:1986mw}. The result is identical to \eqref{eq:4gravAmp}
but with the relevant replacements of $h_{ab}$ by $B_{ab}$ in
$\hat{K}$. Amplitudes with an odd number of $B$ fields trivially
vanish since $\hat{K}$ changes sign under $(a_r,b_r) \leftrightarrow
(c_r,d_r)$. \cite{Gross:1986mw}, where terms involving dilatons were
also studied, showed that the lowest-order contributions to the
effective action can be written exactly as in \eqref{eq:R4} and
\eqref{eq:R4def} but with $R_{abcd}$ everywhere replaced by
\begin{equation} \label{eq:barR}
  \bar{R}_{abcd} = R_{abcd}
  + {\textstyle\half} e^{-\phi/2}\nabla_{[a}H_{b]cd}
  - {\textstyle\quart} g_{[a_{[c}} \nabla_{b]} \nabla_{_{d]}} \phi,
\end{equation}
where $H_{abc}\equiv 3\nabla_{[a}B_{bc]}$ is the field strength
associated with $B_{ab}$. This leads to various new terms such as
$R^2(\nabla H)^2$, $(\nabla H)^4$ and $R^3\nabla\nabla\phi$. The
vanishing of terms involving an odd number of $H$ fields follows from
parity\footnote{This only applies to $t_8t_8$ and
$\epsilon_{10}\epsilon_{10}$ terms. Terms involving a single
$\epsilon_{10}$ are not forbidden, such as the $B\wedge t_8R^4$
term found in IIA \cite{Vafa:1995fj}.}.

\section{The $B^2h^3$ Amplitude and its Low-Energy Expansion} \label{sec:B2h3amp}
Using the light-cone gauge Green-Schwarz formalism, we now calculate
similar one-loop amplitudes but with five rather than four states. The
case of five gravitons was considered in \cite{Richards:2008jg}. Here
we will replace some of these gravitons by \NSNS\ two-forms. It is no
longer true that amplitudes with an odd number of $B$ fields, such as
$Bh^4$, will vanish. However, the non-zero piece will be entirely
contained in the $\epsilon_{10}t_8$ part; the $t_8t_8$ and
$\epsilon_{10}\epsilon_{10}$ terms will still vanish. Since, as
mentioned above, the $\epsilon_{10}t_8$ pieces cannot be seen using
this formalism, we instead choose to focus on the $B^2h^3$ and $B^4h$
cases.

The $B^2h^3$ amplitude proceeds exactly as in the five-graviton case
in \cite{Richards:2008jg}, simply with $h_1$ and $h_2$ replaced by
$B_1$ and $B_2$. Here we only sketch the calculation and refer the
reader to \cite{Richards:2008jg} for details. Let the two-forms have
polarisations $B_1,B_2$ and momenta $k_1,k_2$ respectively, and let
the gravitons have polarisations $h_1,h_2,h_3$ and momenta
$k_3,k_4,k_5$ respectively. The two-forms differ from the gravitons in
that their polarisations are antisymmetric rather than symmetric. Both
the graviton and \NSNS\ two-form vertex operators are given by
\cite{Green:1987sp}
\begin{equation}
  \mathcal{V}_{h,B}(k,z) = \zeta_{ac} (\partial X^a(z) - R^{ab}(z)k^b)
  (\bar\partial X^c(z) - \tilde{R}^{cd}(z)k^d) e^{ik\cdot X(z)},
\end{equation}
where $\zeta_{ab}$ is the polarisation and $R^{ab}(z) \equiv \quart
S(z)^A\gamma_{AB}^{ab}S^B(z)$. Motivated by the usual prescription for
calculating GS amplitudes, explained in \cite{Green:1987mn}, we
consider
\begin{equation}
  A_{B^2h^3} = \int_{\cal F}\!d^2\tau\dV\int d^{10}p\,\tr
  \left(\mathcal{V}_B(k_1,\rho_1)\mathcal{V}_B(k_2,\rho_2)
  \mathcal{V}_h(k_3,\rho_3)\cdots
  w^{L_0}\bar{w}^{\tilde L_0}\right),
\end{equation}
where $v_r = \ln\rho_r/2\pi i$, $\tau = \ln w/2\pi i$, and the trace is
over all $\alpha$, $\tilde\alpha$, $S$ and $\tilde{S}$ modes. The trace
over $S$ vanishes unless there are at least eight $S_0$
zero modes (and similarly for $\tilde{S}$) and so there are only three
types of term to consider: one term containing $R^5\tilde{R}^5$, ten
terms containing $\partial{X}R^4\tilde{R}^5$ or
$R^5\bar\partial{X}\tilde{R}^4$, and twenty-five terms containing
$\partial{X}R^4\bar\partial{X}\tilde{R}^4$.

If we suppress the polarisation tensors and perform the traces and
$p$-integral, then the $R^5\tilde{R}^5$ term can be evaluated as
\begin{align}
  & \dT \dV \prod_{r<s}(\chi_{rs})^{\frac{1}{2}k_r\cdot k_s} k_1^{b_1}
  \cdots k_5^{b_5} k_1^{d_1} \cdots k_5^{d_5} \notag\\
  & \quad \times\Big( t_{10}^{a_1b_1a_2b_2\cdots} + \sum_{r<s}
  \bar{t}_{10}^{\,a_rb_ra_sb_s\cdots}\eta'(v_{rs},\tau) \Big)
  \Big( t_{10}^{c_1d_1c_2d_2\cdots} + \sum_{r<s}
  \bar{t}_{10}^{\,c_rd_rc_sd_s\cdots}\bar\eta'(v_{rs},\tau) \Big),
\end{align}
where, for example, $t_{10}^{\,a_2b_2a_4b_4\cdots}$ is shorthand
for $t_{10}^{\,a_2b_2a_4b_4a_1b_1a_3b_3a_5b_5}$. Here $t_{10}$
and $\bar{t}_{10}$ are ten-index tensors, which can both be
written as sums of $t_8\delta$ tensors, as given in the appendix of
\cite{Richards:2008jg}. The function $\eta'(v_{rs},\tau)$ can be
expressed in terms of Jacobi theta functions via $\pi
i(2\eta'(v,\tau)+1) = -\theta'_1(v,\tau)/\theta_1(v,\tau)$. Similarly,
the $\partial{X}(\rho_1)R^4\tilde{R}^5$ term gives
\begin{align}
  & \dT \dV \prod_{r<s}(\chi_{rs})^{\frac{1}{2}k_r\cdot k_s} k_2^{b_2}
  \cdots k_5^{b_5} k_1^{d_1} \cdots k_5^{d_5} \notag\\
  & \quad \times t_8^{a_2b_2\cdots a_5b_5}
  \Big( \sum_{r\neq 1} k_r^{a_1}\eta(v_{r1},\tau) \Big)
  \Big( t_{10}^{c_1d_1c_2d_2\cdots} + \sum_{r<s}
  \bar{t}_{10}^{\,c_rd_rc_sd_s\cdots}\bar\eta'(v_{rs},\tau) \Big),
\end{align}
where $\eta(v,\tau) = -\eta'(v,\tau)+\frac{\Imagi\,v}{\ImTau}-\half.$
Other $\partial{X}R^4\tilde{R}^5$ terms and the $R^5\bar\partial{X}
\tilde{R}^4$ terms are given by similar expressions. Finally,
evaluating the traces and $p$-integral for, for example,
$\partial{X}(\rho_1)R^4 \bar\partial{X}(\rho_2)\tilde{R}^4$ gives
\begin{align}
  & \dT \dV \prod_{r<s}(\chi_{rs})^{\frac{1}{2}k_r\cdot k_s} k_2^{b_2}
  \cdots k_5^{b_5} k_1^{d_1}k_3^{d_3} \cdots k_5^{d_5} \notag\\
  &\quad\times \left(\sum_{r=2}^5
  k_r^{a_1}\eta(v_{r1},\tau)\sum_{s=1,s\neq 2}^5
  k_s^{c_2}\bar\eta(v_{s2},\tau)
  - 2\delta^{a_1c_2}\hat\Omega(v_{12},\tau)
  \right)t_8^{a_2b_2a_3b_3\cdots}t_8^{c_1d_1c_3d_3\cdots},
\end{align}
where $\hat\Omega(v,\tau)=-1/(2\pi\ImTau)$. Again there are similar
expressions for the other $\partial{X}R^4\bar\partial{X}\tilde{R}^4$
terms.

Using various identities given in the appendix of
\cite{Richards:2008jg}, both $t_{10}$ and $\bar{t}_{10}$ can be
eliminated in favour of $t_8$ tensors. This allows the full amplitude
to be packaged together as
\begin{align} \label{eq:amp1}
  A_{B^2h^3,t_8t_8} &=
  B^1_{a_1c_1}B^2_{a_2c_2}h^1_{a_3c_3}h^2_{a_4c_4}h^3_{a_5c_5}
  \dT \dV \chiprod \notag\\
  & \qquad\times \left( \sum_{r<s}\eta(v_{rs},\tau)A_{rs}
  \sum_{r<s}\bar\eta(v_{rs},\tau)\bar{A}_{rs}
  + \sum_{r<s}\hat\Omega(v_{rs},\tau)C_{rs} \right),
\end{align}
where the indices on $A_{rs}$, $\bar{A}_{rs}$ and $C_{rs}$ have been
suppressed for brevity,
\begin{align} \label{eq:A12}
  A_{12} &= k_1^{a_2}(k_1+k_2)^{b}k_3^{b_3}k_4^{b_4}k_5^{b_5} t_8^{a_1ba_3b_3a_4b_4a_5b_5} \notag\\
  & \quad -k_2^{a_1}(k_1+k_2)^{b}k_3^{b_3}k_4^{b_4}k_5^{b_5} t_8^{a_2ba_3b_3a_4b_4a_5b_5} \notag\\
  & \quad -\delta^{a_1a_2}k_1^{b_1}k_2^{b_2}k_3^{b_3}k_4^{b_4}k_5^{b_5} t_8^{b_1b_2a_3b_3a_4b_4a_5b_5} \notag\\
  & \quad -k_1\cdot k_2\,k_3^{b_3}k_4^{b_4}k_5^{b_5} t_8^{a_1a_2a_3b_3a_4b_4a_5b_5}
\end{align}
and
\begin{align} \label{eq:B12}
  C_{12} = -4\,\delta^{a_1c_2}
  k_2^{b_2}k_3^{b_3}k_4^{b_4}k_5^{b_5} k_1^{d_1}k_3^{d_3}k_4^{d_4}k_5^{d_5}
  t_8^{a_2b_2a_3b_3a_4b_4a_5b_5}t_8^{c_1d_1c_3d_3c_4d_4c_5d_5}.
\end{align}
$\bar{A}_{12}$ is the same as $A_{12}$ but with $c_r$ replacing
$a_r$. The other $A_{rs}$ and $C_{rs}$ are similar but with the
relevant permutations of the momenta and polarisation
indices. The amplitude contains massless poles in $k_r\cdot k_s$ which
originate from the $v_{rs}$ integral over $|\eta(v_{rs},\tau)|^2$.

There is no known way of explicitly evaluating the integrals over
$v_r$ and $\tau$. However, it is still possible to expand them for
small values of $\alpha'$. Since we are only interested in the
expansion to lowest order, there are only two types of integral that
must be considered\footnote{The other possible integrals all vanish at
lowest order in $\alpha'$.},
\begin{align}
  K &= \dT \dVdash \chiproddash \hat\Omega_{rs}, \notag \\
  I_{rs} &= \dT \dVdash \chiproddash |\eta_{rs}|^2,
\end{align}
where $\hat\Omega_{rs}\equiv\hat\Omega(v_{rs},\tau)$ and similarly for
$\eta_{rs}$. Since $\hat\Omega_{rs}$ is independent of $v_{rs}$,
we write $K$ without any subscripts. Using the definition of
$\hat\Omega_{rs}$, the lowest-order expansion of $K$ is easily found
by setting $k_{r'}\cdot k_{s'}$ to zero for all $r'$, $s'$,
\begin{equation}
  K|_{\alpha'^0} = -\frac{1}{2\pi}\int_{\cal F}\!\frac{d^2\tau}{(\ImTau)^2}
  = -\frac{1}{6}.
\end{equation}
At lowest order the expansion of $I_{rs}$ is a pole term and so it
cannot be studied simply by setting $k_{r'}\cdot k_{s'}=0$. However,
it is clear that the pole originates from the corner of the
integration region where $v_r\to v_s$. Then the pole can be extracted
by writing $v_{rs}=|v|e^{i\theta}$ and integrating over a small circle
around $v_{rs}=0$. From the small $v$ behaviour of $\chi_{rs}$ and
$\eta_{rs}$,
\begin{equation}
  \chi(v,\tau) \sim 2\pi|v|, \qquad
  \eta(v,\tau) \sim -\frac{i}{2\pi v},
\end{equation}
it can easily be shown that in the small $k_r\cdot k_s$ limit,
\begin{equation}
  I_{rs} \sim \frac{1}{k_r\cdot k_s} \cdot \frac{1}{\pi} \dT
  \int_{\cal T}\prod_{r'=2}^4 d^2v_{r'}
  {\prod_{\substack{r'<s' \\ 1 \to 2\phantom{'}}}}'
  (\chi_{r's'})^{\half k_r'\cdot k_s'},
\end{equation}
where the prime on the product indicates that $(r',s')=(1,2)$ is not
to be included, and $1 \to 2$ means that $v_{1'}$ is to be replaced by
$v_{2'}$ everywhere within the product. The $\tau$ and $v_{r'}$ integrals are
exactly those that appear in the four-graviton amplitude
\eqref{eq:4gravAmp}, which of course must be the case from unitarity,
and so their low-energy expansion begins with $\pi/3$ as in
\eqref{eq:4gravexp}, giving
\begin{equation}
  I_{rs}|_{\alpha'^{-1}} = \frac{1}{6\alpha'k_r\cdot k_s},
\end{equation}
where $\alpha'$ has been reinstated using $2\alpha'=1$. So the
lowest-order expansion of the amplitude \eqref{eq:amp1} is given by
\begin{equation} \label{eq:amp2}
  A_{B^2h^3,t_8t_8}|_{\alpha'^4} = \frac{2^4\alpha'^4}{6}
  B^1_{a_1c_1}B^2_{a_2c_2}h^1_{a_3c_3}h^2_{a_4c_4}h^3_{a_5c_5}
  \sum_{r<s}\left( \frac{2}{k_r\cdot k_s}|A_{rs}|^2 - C_{rs}
  \right),
\end{equation}
where again we have reinstated $\alpha'$.

Unlike the five-graviton version in \cite{Richards:2008jg}, various
terms within the $|A_{12}|^2$, $|A_{1n}|^2$ and $|A_{2n}|^2$ parts,
with $n=3,4,5$, vanish due to identities such as
$k_1^ak_1^bB_{2ab}^{}=0$ and $B_1^{ab}h_{1ab}^{}=0$. For example, of
the twenty-one terms in $|A_{rs}|^2$, only fifteen survive in the
$|A_{12}|^2$ part,
\begin{align} \label{eq:BBterms}
  \frac{1}{k_1\cdot k_2}\Big(
  & -2k_{1e}k_1^{b_1}k_1^{d_1}{B_1^{a_1}}_f k_2^fB_2^{ec_1}
    -2k_{1e}k_1^{b_1}{B_1^{a_1}}_f k_2^fk_2^{d_1}B_2^{ec_1}
    -2k_{1e}k_1^{b_1}k_1^{c_1}{B_1^{a_1}}_f k_2^{d_1}B_2^{ef} \notag\\
  & -2k_{1e}k_1^{d_1}{B_1^{a_1}}_f k_2^fk_2^{b_1}B_2^{ec_1}
    -2k_{1e}{B_1^{a_1}}_f k_2^fk_2^{b_1}k_2^{d_1}B_2^{ec_1}
    -2k_{1e}k_1^{c_1}{B_1^{a_1}}_f k_2^{b_1}k_2^{d_1}B_2^{ef} \notag\\
  & +2k_1^{b_1}k_1^{c_1}B_{1ef} k_2^{d_1}k_2^{e}B_2^{a_1f}
    +2k_1^{c_1}B_{1ef} k_2^{b_1}k_2^{d_1}k_2^eB_2^{a_1f}
    + k_1^{a_1}k_1^{c_1}B_{1ef} k_2^{b_1}k_2^{d_1}B_2^{ef} \Big) \notag\\
  +\Big(
  & -2k_{1e}k_1^{b_1}B_1^{a_1c_1} B_2^{ed_1}
    -2k_{1e}B_1^{a_1c_1} k_2^{b_1}B_2^{ed_1}
    +2k_1^{b_1}{B_{1e}}^{c_1} k_2^eB_2^{a_1d_1} \notag\\
  & +2{B_{1e}}^{c_1} k_2^ek_2^{b_1}B_2^{a_1d_1}
    +2k_1^{a_1}{B_{1e}}^{c_1} k_2^{b_1}B_2^{ed_1}
    +k_{1e}B_1^{a_1c_1} k_2^eB_2^{b_1d_1} \Big),
\end{align}
which is all multiplied by $\frac{2^4\alpha'^4}{3}
t_{8\,a_1b_1\cdots}t_{8\,c_1d_1\cdots}k_3^{a_2}k_3^{c_2}h_1^{b_2d_2}
k_4^{a_3}k_4^{c_3}h_2^{b_3d_3}k_5^{a_4}k_5^{c_4}h_3^{b_4d_4}$.
Similarly, for the $|A_{13}|^2$ part only seventeen terms remain,
\begin{align} \label{eq:Bhterms}
  \frac{1}{k_1\cdot k_3}\Big(
  & +k_{1e}k_{1f}k_1^{b_1}k_1^{d_1}B_1^{a_1c_1} h_1^{ef}
    +2k_{1e}k_{1f}k_1^{b_1}B_1^{a_1c_1} k_3^{d_1}h_1^{ef}
    -2k_{1e}k_1^{b_1}k_1^{d_1}{B_1^{a_1}}_f k_3^fh_1^{ec_1}\notag\\
  & -2k_{1e}k_1^{b_1}{B_1^{a_1}}_f k_3^fk_3^{d_1}h_1^{ec_1}
    -2k_{1e}k_1^{b_1}k_1^{c_1}{B_1^{a_1}}_f k_3^{d_1}h_1^{ef}
    + k_{1e}k_{1f}B_1^{a_1c_1} k_3^{b_1}k_3^{d_1}h_1^{ef}\notag\\
  & -2k_{1e}k_1^{d_1}{B_1^{a_1}}_f k_3^fk_3^{b_1}h_1^{ec_1}
    -2k_{1e}{B_1^{a_1}}_f k_3^fk_3^{b_1}k_3^{d_1}h_1^{ec_1}
    -2k_{1e}k_1^{c_1}{B_1^{a_1}}_f k_3^{b_1}k_3^{d_1}h_1^{ef} \notag\\
  & +2k_1^{b_1}k_1^{c_1}B_{1ef} k_3^{d_1}k_3^eh_1^{a_1f}
    +2k_1^{c_1}B_{1ef} k_3^{b_1}k_3^{d_1}k_3^eh_1^{a_1f} \Big) \notag\\
  + \Big(
  & -2k_{1e}k_1^{b_1}B_1^{a_1c_1} h_1^{ed_1}
    -2k_{1e}B_1^{a_1c_1} k_3^{b_1}h_1^{ed_1}
    +2k_1^{b_1}{B_{1e}}^{c_1} k_3^eh_1^{a_1d_1}\notag\\
  & +2{B_{1e}}^{c_1} k_3^ek_3^{b_1}h_1^{a_1d_1}
    +2k_1^{a_1}{B_{1e}}^{c_1} k_3^{b_1}h_1^{ed_1}
    +k_{1e}B_1^{a_1c_1} k_3^eh_1^{b_1d_1} \Big),
\end{align}
which is all multiplied by $\frac{2^4\alpha'^4}{3}
t_{8\,a_1b_1\cdots}t_{8\,c_1d_1\cdots} k_2^{a_2}k_2^{c_2}B_2^{b_2d_2}
k_4^{a_3}k_4^{c_3}h_2^{b_3d_3} k_5^{a_4}k_5^{c_4}h_3^{b_4d_4}$. These
expressions will be important for matching with the field theory
diagrams in the next section.

\section{Consequences for the Effective Action} \label{sec:EA}
The expanded amplitude in \eqref{eq:amp2} can now be compared with the
same amplitude calculated from known quartic terms in the effective
action, such as $R^4$ and $R^2(\nabla H)^2$. These terms, however,
will not account for the full amplitude, and the remainder will
require new $H^2R^3$ terms.

At lowest order the effective action consists of the usual
supergravity terms, which in Einstein frame are given by
\begin{equation} \label{eq:SUGRAEffAct}
  S_{\alpha'^0} = \int d^{10}x\sqrt{-g}\,( R -{\textstyle\frac{1}{12}}
  e^{-\phi}H^2 -{\textstyle\half}(\partial\phi)^2 ).
\end{equation}
Einstein frame is used to avoid mixing between the dilaton and
graviton propagators. It is important to include the dilaton since
there are diagrams where a dilaton propagates as an intermediate
particle. As mentioned in section \ref{sec:4ptAmps}, the first
correction to $S_{\alpha'^0}$ occurs at order $\alpha'^3$. For our
purposes, only the one-loop correction is relevant,
\begin{equation} \label{eq:GrossSloanEA}
  S_{\alpha'^3,{\rm 1-loop}} = \frac{2\pi^2}{3}\alpha'^3\int d^{10}x
  \sqrt{-g}\, e^{\phi/2}\, t_8^{a_1b_1\cdots}t_8^{c_1d_1\cdots}
  \bar{R}_{a_1b_1c_1d_1}\cdots\bar{R}_{a_4b_4c_4d_4},
\end{equation}
where $\bar{R}_{abcd}$ is given in \eqref{eq:barR}. In particular, for
matching with the $B^2h^3$ amplitude, only the $t_8t_8R^4$,
$t_8t_8R^2(\nabla H)^2$ and $t_8t_8R^3(\nabla\nabla\phi)$ terms are
required.

\subsection{Expansion of Various Tensors}
Before we can expand the terms in $S_{\alpha'^0}$ and $S_{\alpha'^3}$,
we need the expansions of the various fields and tensors
involved. Consider a small fluctuation of the metric about the
Minkowski metric, $g_{ab} = \eta_{ab} + \kappa h_{ab}$, where $\kappa$
is presumed small. In subsequent expressions we will drop factors of
$\kappa$ since they can easily be reinstated. The expansions of the
Riemann tensor, the Ricci scalar and the $t_8$ tensor were given in
\cite{Richards:2008jg} and we refer the reader there for details.

At most we require the expansion of the Riemann tensor, $R_{abcd}$, to
second order in $h$. As explained in \cite{Richards:2008jg}, since we
are only concerned with five-point amplitudes, a Riemann tensor
expanded to second order is guaranteed to be multiplied by a tensor
which is antisymmetric in $a\leftrightarrow b$ and $c\leftrightarrow
d$, and symmetric in $(a,\,b)\leftrightarrow(c,\,d)$. With this
understanding, the expansion to second order simplifies to
\begin{equation} \label{eq:RiemannExp}
  R_{abcd} = 2\partial_a \partial_c h_{bd}
    + \partial_a h_c^{\phantom{c}e} \partial_d h_{be} +
    \partial_a h_c^{\phantom{c}e} \partial_b h_{de}
     - 2\partial^e h_{ac} \partial_b h_{de}
     + {\textstyle\half} \partial^e h_{ac} \partial_e h_{bd}.
\end{equation}
The expansion of the Ricci scalar begins
\begin{align}
  R &= \Box h - \partial_a\partial_b h^{ab} \notag\\
  &\quad - h^{ab}(\Box h_{ab} + \partial_a\partial_bh
    - 2\partial_a\partial^ch_{bc}) \notag\\
  &\quad - {\textstyle\frac{3}{4}}\partial_ah_{bc}\partial^ah^{bc}
    + {\textstyle\half}\partial_ah_{bc}\partial^bh^{ac}
    + \partial^ah_{ab}\partial_ch^{bc} - \partial^ah_{ab}\partial^bh
    + {\textstyle\quart}\partial^ah\partial_ah,
\end{align}
where $h={h^a}_a$. Although we actually need the expansion to third
order in $h$, we do not need the explicit expression and so there is
no need to write it here.

Since the $t_8$ tensor is formed from products of the metric, it is
important to also consider its expansion. As for the Riemann tensor,
since whenever $t_8$ is expanded to first order it is always
multiplied by tensors which are symmetric under, for example,
$(a,\,b)\leftrightarrow(c,\,d)$, the expansion reduces to
\begin{equation} \label{eq:t8exp}
  t_8^{abcdefgh} = \underline{t}_8^{abcdefgh} -2 (
  {h_i}^a\underline{t}_8^{ibcdefgh} +
  {h_i}^b\underline{t}_8^{aicdefgh} ),
\end{equation}
where $t_8$ is formed out of products of the curved metric, $g$, and
$\underline{t}_8$ is the equivalent expression formed out of the
Minkowski metric, $\eta$.

The expansion of $H^2$ is achieved using
\begin{equation}
  H^2 \equiv g^{ad}g^{be}g^{cf}H_{abc}H_{def} \equiv
  3^2g^{ad}g^{be}g^{cf}\nabla_{[a}B_{bc]}\nabla_{[d}B_{ef]}
\end{equation}
and remembering that it is also necessary to expand the covariant
derivatives. It is readily found that, up to first order in $h$,
\begin{align} \label{eq:H2exp}
  2H^2 &= \partial_aB_{bc} (\partial^aB^{bc}+2\partial^bB^{ca}) \notag\\
  &\qquad - h^{ad}(\partial_aB_{bc}\partial_dB^{bc}
                  -4\partial_aB_{bc}\partial^b{B_d}^c
                  +2\partial_bB_{ac}\partial^b{B_d}^c
                  -2\partial_bB_{ac}\partial^c{B_d}^b).
\end{align}

Since we need the expansion of $S_{\alpha'^3}$ up to terms involving
five fields, we require the expansion of $\nabla_{[a}H_{b]cd}$ up to
first order in $h$. The zeroth order contribution is trivial and the
first order terms originate from expanding the Christoffel symbols
within the derivative. When $\nabla_{[a}H_{b]cd}$ is expanded to first
order it is assured of being multiplied by an expression which is
manifestly antisymmetric in $a \leftrightarrow b$, antisymmetric in $c
\leftrightarrow d$, and {\it antisymmetric} in
$(a,b)\leftrightarrow(c,d)$\footnote{This antisymmetry follows since,
from the Bianchi identity, $\nabla_{[a}H_{b]cd}=-\nabla_{[c}H_{d]ab}$.}.
With the understanding that $\nabla_{[a}H_{b]cd}$ is multiplied by a
tensor with such symmetries, the expansion simplifies and, up to first
order in $h$, can be written as
\begin{align} \label{eq:DHexp}
  {\textstyle\half} \nabla_{[a}H_{b]cd} &= \partial_a \partial_c B_{db}
     + \partial_a h_c^{\phantom{c}e} \partial_d B_{be}
     +  \partial_a h_c^{\phantom{c}e} \partial_b B_{ed} \notag\\
   &\qquad - \partial^e h_{ac} \partial_b B_{ed}
     - \partial_a h_c^{\phantom{c}e} \partial_e B_{bd}
     + {\textstyle\half} \partial^e h_{ac} \partial_e B_{bd},
\end{align}
which, although similar to the expansion of the Riemann tensor,
differs due to the different symmetries of $h_{ab}$ and $B_{ab}$, and
the missing factor of two in the zeroth order term.

\subsection{Propagators}
From \eqref{eq:SUGRAEffAct} we can derive the propagators for the
graviton, the \NSNS\ two-form and the dilaton. Since the dilaton is a
scalar, its propagator is simply $D = 1/k^2$. For the graviton we
consider the Einstein-Hilbert term which, after dropping total
derivatives, is given up to second order by,
\begin{equation}
  S_{\alpha^0,R} = \quart\int d^{10}x\, (\partial_ah_{bc}\partial^ah^{bc} -
  \partial_ah\partial^ah +2\partial_ah\partial_bh^{ab} -2
  \partial_ah_{bc}\partial^bh^{ac}),
\end{equation}
which is invariant under the gauge transformation
\begin{equation}
  h_{ab} \to h_{ab} + \partial_a\zeta_b + \partial_b\zeta_a,
\end{equation}
where $\zeta_a$ is an arbitrary one-form field. After fixing the gauge
invariance using the de Donder gauge, $\partial^ah_{ab} =
{\textstyle\half}\partial_bh$, the graviton propagator can easily be
shown to be
\begin{equation} \label{eq:gravprop}
  D_{ab,cd} = \frac{\eta_{ac}\eta_{bd} + \eta_{ad}\eta_{bc} -
  \quart\eta_{ab}\eta_{cd}}{k^2}.
\end{equation}

To find the propagator for the two-form we first fix the gauge
invariance, $B_{ab}\to B_{ab}+\nabla_{[a}\zeta_{b]}$, by adding the
gauge fixing term $\lambda \partial_aB^{ac} \partial_b{B^b}_c$ to the
action. Then, after removing a total derivative, the part of
\eqref{eq:SUGRAEffAct} quadratic in $B_{ab}$ becomes
\begin{equation}
  S_{\alpha'^0,H^2} = 
    -\frac{1}{24}\int d^{10}x
    (\partial_aB_{bc}\partial^aB^{bc}
    -2B_{bc}\partial_a\partial^bB^{ca}
    +24\lambda B^{ac}\partial_a\partial_b{B^b}_c ).
\end{equation}
Now we write the integrand as $B_{ab}V^{abcd}B_{cd}$ and try to invert
$V^{abcd}$, by which we mean solve $D'_{ab,cd}{V^{cd}}_{ef} =
\eta_{[a|e|}\eta_{b]f}$. This can only be achieved for the particular
choice $\lambda=-\frac{1}{12}$, after which we find
\begin{equation} \label{eq:Bprop}
  D'_{ab,cd} = \frac{\eta_{ac}\eta_{bd}-\eta_{ad}\eta_{bc}}{k^2}.
\end{equation}

\subsection{Evaluation of Diagrams}
\begin{figure}
\begin{center}
\includegraphics[scale=0.5]{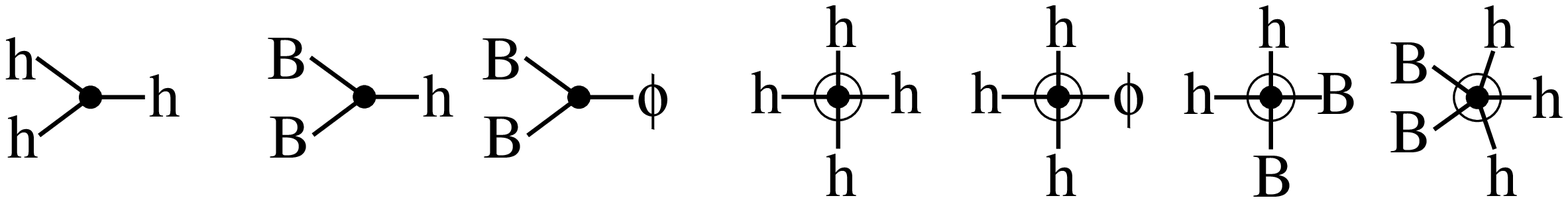}
\caption{Field theory vertices relevant for the $B^2h^3$
  amplitude. From left to right: a three-vertex from $R$, two
  three-vertices from $e^{-\phi}H^2$, and three four-vertices and a
  five-vertex from $\bar{R}^4$.}
\label{fig:EffActVertices}
\end{center}
\end{figure}

The expansions of $S_{\alpha'^0}$ and $S_{\alpha'^3}$ lead to several
three-, four- and five-vertices which appear in $B^2h^3$
diagrams. Firstly, the Einstein-Hilbert term contains the usual
three-graviton vertex. Further, the kinetic term for the \NSNS\
two-form, $e^{-\phi}H^2 = H^2 - \phi H^2 + \ldots$,  gives a $BBh$
three-vertex from the expansion of the first term, and a $BB\phi$
three-vertex from the second term. The quartic one-loop term,
$S_{\alpha'^3}$, leads to three relevant four-vertices: a
four-graviton vertex, a three-graviton and one-dilaton vertex, and a
two-graviton and two-$B$-field vertex. Finally, the $R^2(\nabla H)^2$ term
in $S_{\alpha'^3}$ generates a $BBhhh$ five-vertex. These vertices are
shown in figure \ref{fig:EffActVertices} where a vertex surrounded by
a circle originates from the one-loop $S_{\alpha'^3}$ term, whereas a
vertex without a circle originates from the tree-level $S_{\alpha'^0}$
term.

\begin{figure}
\begin{center}
\includegraphics[scale=0.5]{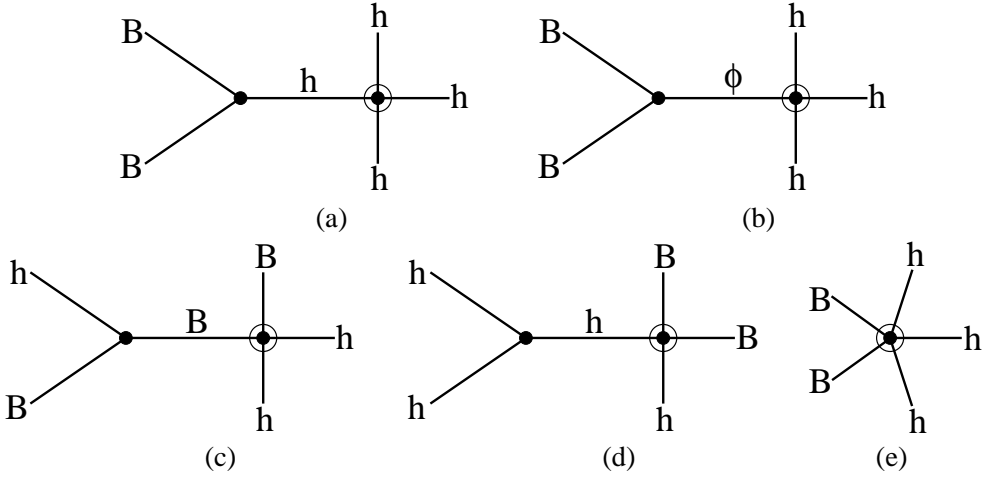}
\caption{Field theory diagrams contributing to the $B^2h^3$
  amplitude.}
\label{fig:EffActDiags}
\end{center}
\end{figure}

Unlike the five-graviton amplitude in \cite{Richards:2008jg} where
there were only two field theory diagrams to calculate, there are now five
separate diagrams to evaluate (figure \ref{fig:EffActDiags}). Diagrams
(a) and (b) both contain internal states carrying momenta $\sqrt{-2k_1
\cdot k_2}$, the first with an intermediate graviton and the second
with an  intermediate dilaton. Diagrams (c) and (d) are similar: the
first contains an intermediate $B$-field carrying momenta $\sqrt{-2k_r
\cdot k_s}$ and the second contains an intermediate graviton carrying
momenta $\sqrt{-2k_s\cdot k_{s'}}$, where $r\in\{1,2\}$ and
$s,s'\in\{3,4,5\}$. Diagram (e) is a contact diagram formed from a
single five-vertex. All these diagrams must be evaluated and
subtracted from the $B^2h^3$ amplitude, before the remaining terms can
be covariantised.

Diagram (a) consists of a $BBh$ three-vertex connected via a graviton
propagator to a four-graviton one-loop vertex and can be evaluated as
follows. The three-vertex is found from \eqref{eq:H2exp} and first
must be contracted into the two on-shell $B$-fields, $B_1$ and
$B_2$. The two free indices are then contracted with the graviton
propagator \eqref{eq:gravprop}, before multiplying by the
four-graviton vertex derived from $t_8t_8R^4$. Finally, the remaining
external legs contract into the three gravitons, $h_3$, $h_4$ and
$h_5$. After some work, the result simplifies to
\begin{align} \label{eq:B2h3DiagramA}
  \parbox{12mm}
  {\includegraphics[scale=0.3]{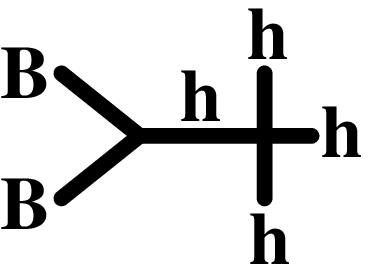}}
  &= t_{8\,a_1b_1\cdots}t_{8\,c_1d_1\cdots} k_3^{a_2}k_3^{c_2}h_1^{b_2d_2}
     \, k_4^{a_3}k_4^{c_3}h_2^{b_3d_3} \, k_5^{a_4}k_5^{c_4}h_3^{b_4d_4} \notag\\
  & \qquad\times\raisebox{0pt}[0pt][0pt]{\Bigg(} \frac{1}{k_1\cdot k_2} \Big(
    -2k_{1e}k_1^{b_1}k_1^{d_1}{B_1^{a_1}}_f k_2^fB_2^{ec_1}
    -2k_{1e}k_1^{b_1}{B_1^{a_1}}_f k_2^fk_2^{d_1}B_2^{ec_1}\notag\\
  & \qquad\qquad\qquad\quad~\, -2k_{1e}k_1^{b_1}k_1^{c_1}{B_1^{a_1}}_f k_2^{d_1}B_2^{ef}
    -2k_{1e}k_1^{d_1}{B_1^{a_1}}_f k_2^fk_2^{b_1}B_2^{ec_1} \notag\\
  & \qquad\qquad\qquad\quad~\, -2k_{1e}{B_1^{a_1}}_f k_2^fk_2^{b_1}k_2^{d_1}B_2^{ec_1}
    -2k_{1e}k_1^{c_1}{B_1^{a_1}}_f k_2^{b_1}k_2^{d_1}B_2^{ef} \notag\\
  & \qquad\qquad\qquad\quad~\, +2k_1^{b_1}k_1^{c_1}B_{1ef} k_2^{d_1}k_2^{e}B_2^{a_1f}
    +2k_1^{c_1}B_{1ef} k_2^{b_1}k_2^{d_1}k_2^eB_2^{a_1f} \notag\\
  & \qquad\qquad\qquad\quad~\, + k_1^{a_1}k_1^{c_1}B_{1ef} k_2^{b_1}k_2^{d_1}B_2^{ef} \notag\\
  & \qquad\qquad\qquad\quad~\, -\frac{3}{4}(k_1+k_2)^{a_1}(k_1+k_2)^{c_1}k_{1e}B_{1fg}
    k_2^gB_2^{fe}\eta^{b_1d_1}\Big) \notag\\
  & \qquad\qquad~ + (k_1+k_2)^{a_1}(k_1+k_2)^{c_1}\Big( 2B_{1e}^{\phantom{1e}(d_1} B_2^{b_1)e}
    + \frac{3}{8}B_{1ef} B_2^{ef} \eta^{b_1d_1} \Big) \raisebox{0pt}[0pt]{\Bigg)},
\end{align}
where all terms apart from those in the last line are poles. The
penultimate line is a pole containing an $\eta^{ab}$ factor. Since
there are no such terms in the $B^2h^3$ amplitude, this must cancel
with an equivalent term from another diagram.

Diagram (b) is similar in spirit to diagram (a), but with a dilaton as
the intermediate particle. As such, the simpler dilaton propagator is
used to contract the two vertices. The three-vertex now originates
from the $\phi H^2$ term in $S_{\alpha'^0}$. The diagram is easily
evaluated as
\begin{align} \label{eq:B2h3DiagramB}
  \parbox{12mm}
  {\includegraphics[scale=0.3]{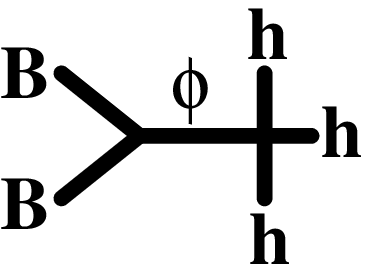}}
  &= t_{8\,a_1b_1\cdots}t_{8\,c_1d_1\cdots} k_3^{a_2}k_3^{c_2}h_1^{b_2d_2}
     \, k_4^{a_3}k_4^{c_3}h_2^{b_3d_3} \, k_5^{a_4}k_5^{c_4}h_3^{b_4d_4} \notag\\
  & \qquad\times (k_1+k_2)^{a_1}(k_1+k_2)^{c_1}\eta^{b_1d_1} \left(
     \frac{3}{4k_1\cdot k_2}k_{1e}B_{1fg}k_2^gB_2^{fe}
    -\frac{3}{8}B_{1ef}B_2^{ef} \right).
\end{align}
The pole term has exactly the correct form to cancel the pole
containing $\eta^{ab}$ in the previous diagram, and the second term
exactly cancels the final term in \eqref{eq:B2h3DiagramA}. So the sum
of diagrams (a) and (b) is given by \eqref{eq:B2h3DiagramA}, but
without the two terms containing $\eta^{ab}$ factors.

Diagram (c) accounts for the poles where a B-field and a graviton are
separated from the other particles. Consider the particular case where
the separated B-field is $B_1$ with momentum $k_1$ and the separated
graviton is $h_1$ with momentum $k_3$. We start from the $BBh$
three-vertex given by \eqref{eq:H2exp} and contract $B_1$ and $h_1$
into two of the legs. The remaining leg is now connected by the
two-form propagator \eqref{eq:Bprop} to a $B^2h^2$ four-vertex from
the expansion of $t_8t_8\bar{R}^4$. Finally, the remaining three legs
are contracted into $B_2$, $h_2$ and $h_3$. After simplifying the
result we obtain
\begin{align} \label{eq:B2h3DiagramC}
  \parbox{12mm}
  {\includegraphics[scale=0.3]{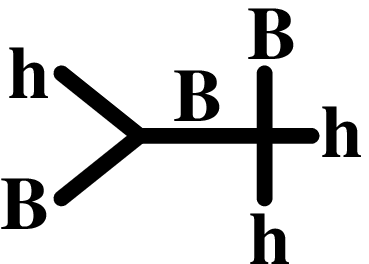}}
  &= t_{8\,a_1b_1\cdots}t_{8\,c_1d_1\cdots} k_2^{a_2}k_2^{c_2}B_2^{b_2d_2}
     \, k_4^{a_3}k_4^{c_3}h_2^{b_3d_3} \, k_5^{a_4}k_5^{c_4}h_3^{b_4d_4} \notag\\
  & \qquad\times\raisebox{0pt}[0pt][0pt]{\Bigg(} \frac{1}{k_1\cdot k_3} \Big(
    +k_{1e}k_{1f}k_1^{b_1}k_1^{d_1}B_1^{a_1c_1} h_1^{ef} 
    +2k_{1e}k_{1f}k_1^{b_1}B_1^{a_1c_1} k_3^{d_1}h_1^{ef} \notag\\
  & \qquad\qquad\qquad\quad~\, -2k_{1e}k_1^{b_1}k_1^{d_1}{B_1^{a_1}}_f k_3^fh_1^{ec_1}
    -2k_{1e}k_1^{b_1}{B_1^{a_1}}_f k_3^fk_3^{d_1}h_1^{ec_1} \notag\\
  & \qquad\qquad\qquad\quad~\, -2k_{1e}k_1^{b_1}k_1^{c_1}{B_1^{a_1}}_f k_3^{d_1}h_1^{ef}
    + k_{1e}k_{1f}B_1^{a_1c_1} k_3^{b_1}k_3^{d_1}h_1^{ef} \notag\\
  & \qquad\qquad\qquad\quad~\, -2k_{1e}k_1^{d_1}{B_1^{a_1}}_f k_3^fk_3^{b_1}h_1^{ec_1}
    -2k_{1e}{B_1^{a_1}}_f k_3^fk_3^{b_1}k_3^{d_1}h_1^{ec_1} \notag\\
  & \qquad\qquad\qquad\quad~\, -2k_{1e}k_1^{c_1}{B_1^{a_1}}_f k_3^{b_1}k_3^{d_1}h_1^{ef}
    +2k_1^{b_1}k_1^{c_1}B_{1ef} k_3^{d_1}k_3^eh_1^{a_1f} \notag\\
  & \qquad\qquad\qquad\quad~\, +2k_1^{c_1}B_{1ef} k_3^{b_1}k_3^{d_1}k_3^eh_1^{a_1f}\Big) \notag\\
  & \qquad\qquad~ + 2(k_1+k_3)^{a_1}(k_1+k_3)^{c_1} B_{1e}^{\phantom{1e}[d_1} h_1^{b_1]e}
    \raisebox{0pt}[0pt]{\Bigg)}.
\end{align}

Diagram (d) is almost identical to the pole diagram calculated in
\cite{Richards:2008jg}, which itself was derived from a similar
amplitude in \cite{Sannan:1986tz}, the only difference being the
four-vertex. This difference, however, is minimal since $R_{abcd}$ and
$\nabla_{[a}H_{b]cd}$ expanded to lowest order both have the form
$2\partial_a\partial_cX_{db}$, where $X$ is $h$ and $B$
respectively. As such, we can simply take the result in
\cite{Richards:2008jg} and replace the relevant gravitons by
B-fields. Let the two gravitons which are to the left of the
propagator be $h_1$ and $h_2$ with momenta $k_3$ and $k_4$
respectively. Then
\begin{align} \label{eq:B2h3DiagramD}
  \parbox{12mm}
  {\includegraphics[scale=0.3]{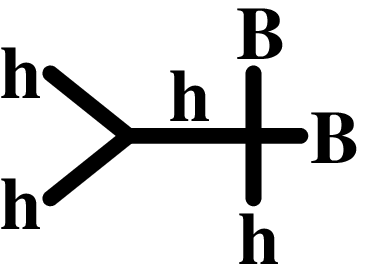}}
  &= t_{8\,a_1b_1\cdots}t_{8\,c_1d_1\cdots} k_1^{a_2}k_1^{c_2}B_1^{b_2d_2}
     \, k_2^{a_3}k_2^{c_3}B_2^{b_3d_3} \, k_5^{a_4}k_5^{c_4}h_3^{b_4d_4} \notag\\
  & \qquad\times\raisebox{0pt}[0pt][0pt]{\Bigg(} \frac{1}{k_3\cdot k_4} \Big(
    +k_{3e}k_{3f}k_3^{b_1}k_3^{d_1}h_1^{a_1c_1} h_2^{ef} 
    +2k_{3e}k_{3f}k_3^{b_1}h_1^{a_1c_1} k_4^{d_1}h_2^{ef} \notag\\
  & \qquad\qquad\qquad\quad~\, -2k_{3e}k_3^{b_1}k_3^{d_1}{h_1^{a_1}}_f k_4^fh_2^{ec_1}
    -2k_{3e}k_3^{b_1}{h_1^{a_1}}_f k_4^fk_4^{d_1}h_2^{ec_1} \notag\\
  & \qquad\qquad\qquad\quad~\, -2k_{3e}k_3^{b_1}k_3^{c_1}{h_1^{a_1}}_f k_4^{d_1}h_2^{ef}
    + k_{3e}k_{3f}h_1^{a_1c_1} k_4^{b_1}k_4^{d_1}h_2^{ef} \notag\\
  & \qquad\qquad\qquad\quad~\, -2k_{3e}k_3^{d_1}{h_1^{a_1}}_f k_4^fk_4^{b_1}h_2^{ec_1}
    -2k_{3e}{h_1^{a_1}}_f k_4^fk_4^{b_1}k_4^{d_1}h_2^{ec_1} \notag\\
  & \qquad\qquad\qquad\quad~\, -2k_{3e}k_3^{c_1}{h_1^{a_1}}_f k_4^{b_1}k_4^{d_1}h_2^{ef}
    +k_3^{b_1}k_3^{d_1}h_{1ef} k_4^ek_4^fh_2^{a_1c_1} \notag\\
  & \qquad\qquad\qquad\quad~\, +2k_3^{b_1}h_{1ef} k_4^ek_4^fk_4^{d_1}h_2^{a_1c_1}
    +2k_3^{b_1}k_3^{c_1}h_{1ef} k_4^{d_1}k_4^eh_2^{a_1f} \notag\\
  & \qquad\qquad\qquad\quad~\, +h_{1ef} k_4^{b_1}k_4^{d_1}k_4^ek_4^fh_2^{a_1c_1}
    +2k_3^{c_1}h_{1ef} k_4^{b_1}k_4^{d_1}k_4^eh_2^{a_1f} \notag\\
  & \qquad\qquad\qquad\quad~\, +k_3^{a_1}k_3^{c_1}h_{1ef} k_4^{b_1}k_4^{d_1}h_2^{ef}\Big) \notag\\
  & \qquad\qquad~ + 2(k_3+k_4)^{a_1}(k_3+k_4)^{c_1} h_{1e}^{\phantom{1e}(d_1} h_2^{b_1)e}
    \raisebox{0pt}[0pt]{\Bigg)}.
\end{align}

Finally we calculate diagram (e), which involves expanding the
$R^2(\nabla H)^2$ term from the one-loop $S_{\alpha'^3}$ action,
\begin{equation}
  \int d^{10}x\sqrt{-g}\,t_8^{a_1b_1\cdots} t_8^{c_1d_1\cdots}
  \nabla_{[a_1}H_{b_1]c_1d_1}\nabla_{[a_2}H_{b_2]c_2d_2}
  R_{a_3b_3c_3d_3}R_{a_4b_4c_4d_4},
\end{equation}
to third order in $h$ and contracting all legs into the on-shell
external states. Since at lowest order $S_{\alpha'^3}$ contains two
gravitons, it is necessary to expand to one order higher than
leading-order. The fifth graviton cannot originate from
$\sqrt{-g}\approx 1+\half {h^a}_a$ since the external gravitons are
traceless. However, it can originate either from a $t_8$ tensor, from
a Riemann tensor or from a $\nabla_{[a}H_{b]cd}$ factor. After a
slightly involved calculation which uses \eqref{eq:t8exp},
\eqref{eq:RiemannExp} and \eqref{eq:DHexp}, we find
\begin{align} \label{eq:B2h3DiagramE}
  \parbox{8mm}
  {\includegraphics[scale=0.3]{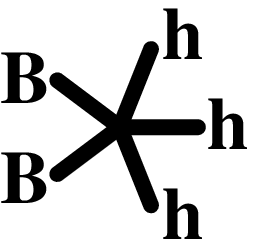}}
  &= 2^5t_{8\,a_1b_1\cdots}t_{8\,c_1d_1\cdots} k_2^{a_2}k_2^{c_2}B_2^{b_2d_2}
     \, k_4^{a_3}k_4^{c_3}h_2^{b_3d_3} \, k_5^{a_4}k_5^{c_4}h_3^{b_4d_4} \notag\\
  & \qquad\times\Big( +k_1^{d_1}{B_1^{b_1}}_e k_3^{a_1}h_1^{c_1e}
    +k_1^{b_1}B_{1e}^{\phantom{1e}d_1} k_3^{a_1}h_1^{c_1e}
    -k_1^{b_1}B_{1e}^{\phantom{1e}d_1} k_3^eh_1^{a_1c_1} \notag\\
  & \qquad\quad~~\, -k_{1e}B_1^{b_1d_1} k_3^{a_1}h_1^{c_1e}
    +{\textstyle\half} k_{1e}B_1^{b_1d_1} k_3^eh_1^{a_1c_1}
    -k_{1e}k_1^{c_1}B_1^{b_1d_1} h_1^{a_1e} \notag\\
  & \qquad\quad~~\, -k_1^{a_1}k_1^{c_1}{B_{1e}}^{d_1} h_1^{b_1e} \Big) \notag\\
  &\quad +2^4t_{8\,a_1b_1\cdots}t_{8\,c_1d_1\cdots} k_1^{a_2}k_1^{c_2}B_1^{b_2d_2}
     \, k_2^{a_3}k_2^{c_3}B_2^{b_3d_3} \, k_5^{a_4}k_5^{c_4}h_3^{b_4d_4} \notag\\
  & \qquad\quad\times\Big( +k_3^{d_1}{h_1^{b_1}}_e k_4^{a_1}h_2^{c_1e}
    +k_3^{b_1}h_{1e}^{\phantom{1e}d_1} k_4^{a_1}h_2^{c_1e}
    -k_3^{b_1}h_{1e}^{\phantom{1e}d_1} k_4^eh_2^{a_1c_1} \notag\\
  & \qquad\qquad~~\, -k_{3e}h_1^{b_1d_1} k_4^{a_1}h_2^{c_1e}
    +{\textstyle\half} k_{3e}h_1^{b_1d_1} k_4^eh_2^{a_1c_1}
    -k_{3e}k_3^{c_1}h_1^{b_1d_1} h_2^{a_1e} \notag\\
  & \qquad\qquad~~\, -k_3^{a_1}k_3^{c_1}{h_{1e}}^{d_1} h_2^{b_1e}
    -{h_1^{b_1}}_e k_4^{a_1}k_4^{c_1}h_2^{d_1e}
    +{h_1^{d_1}}_e k_4^ek_4^{a_1}h_2^{b_1c_1} \Big) \notag\\
  & \quad\mbox{+ all permutations of ($B_1$, $B_2$) and of ($h_1$, $h_2$, $h_3$)}.
\end{align}

\subsection{New $H^2R^3$ Terms}
All the above diagrams need to be subtracted from \eqref{eq:amp2},
before the remainder can be covariantised to discover new $H^2R^3$
terms. The matching of the pole terms is guaranteed from unitarity and
indeed it is readily seen that the poles in the sum of
\eqref{eq:B2h3DiagramA} and \eqref{eq:B2h3DiagramB} match with the
poles in \eqref{eq:BBterms}, that the poles in \eqref{eq:B2h3DiagramC}
match with the poles in \eqref{eq:Bhterms}, and that the poles in
\eqref{eq:B2h3DiagramD} match with the poles in the $|A_{rs}|^2$ part
of \eqref{eq:amp2} for $r,s=3,4,5$.

This leaves the non-poles. In the amplitude these arise from both the
$|A_{rs}|^2$ and $C_{rs}$ terms. In the field theory they originate
from both the non-pole terms in diagrams (a) to (d) and from the
entirety of the contact diagram. Consider first terms where two
gravitons are singled out. As explained in \cite{Richards:2008jg}, it
is always possible to single out such terms even for the
non-poles. Then the non-poles in \eqref{eq:amp2} can easily be seen to
match the non-poles from \eqref{eq:B2h3DiagramD} and the second half
of \eqref{eq:B2h3DiagramE}. The matching is practically identical to
that for the lowest-order expansion of the five-graviton amplitude in
\cite{Richards:2008jg}.

Next consider terms where a $B$-field and a graviton are singled
out. For concreteness assume $B_1$ and $h_1$ to be the separated
fields. Then we need to compare the final two lines in
\eqref{eq:Bhterms} and the $C_{13}$ term with the final line in
\eqref{eq:B2h3DiagramC} and the first half of
\eqref{eq:B2h3DiagramE}. After doing so, not all terms cancel; the
remainder are given by
\begin{equation} \label{eq:BhLeftOverTerms}
  2t_{8\,a_1b_1\cdots}t_{8\,c_1d_1\cdots}
  k_2^{a_2}k_2^{c_2}B_2^{b_2d_2} k_4^{a_3}k_4^{c_3}h_2^{b_3d_3}
  k_5^{a_4}k_5^{c_4}h_3^{b_4d_4}
  \big( {B_{1e}}^{c_1} k_3^ek_3^{b_1}h_1^{a_1d_1} 
  - {B_{1e}}^{a_1} k_3^{b_1}k_3^{d_1}h_1^{c_1e} \big).
\end{equation}
Such terms are potentially problematic since they cannot be
covariantised into new effective action terms. An attempt to do so would
generate terms of the form $t_8t_8B(\nabla H)R^3$, but such terms
cannot appear since they are not gauge-invariant. So there must be a
mechanism whereby these extra terms cancel against other terms.

Finally consider terms where the two $B$-fields are singled out. This
involves comparing the final two lines in \eqref{eq:BBterms} and the
$C_{12}$ term with the penultimate term in \eqref{eq:B2h3DiagramA};
the contact diagram \eqref{eq:B2h3DiagramE} now makes no
contribution. After cancelling the field theory diagrams, the
remaining terms are given by
\begin{align} \label{eq:BBLeftOverTerms}
  &2t_{8\,a_1b_1\cdots}t_{8\,c_1d_1\cdots}
  k_3^{a_2}k_3^{c_2}h_1^{b_2d_2} k_4^{a_3}k_4^{c_3}h_2^{b_3d_3}
  k_5^{a_4}k_5^{c_4}h_3^{b_4d_4} \notag\\
  &\quad\times\big(
    +{B_{1e}}^{c_1} k_2^ek_2^{b_1}B_2^{a_1d_1}
    -{B_{1e}}^{a_1} k_2^{b_1}k_2^{d_1}B_2^{c_1e}
    -k_{1e}k_1^{b_1}B_1^{a_1c_1} B_2^{ed_1} \notag\\
  &\qquad~~
    -k_1^{b_1}k_1^{d_1}{B_{1e}}^{a_1} B_2^{c_1e}
    -k_{1e}B_1^{a_1c_1} k_2^{b_1}B_2^{ed_1}
    +k_1^{b_1}{B_{1e}}^{c_1} k_2^eB_2^{a_1d_1} \notag\\
  &\qquad~~
    +k_1^{a_1}{B_{1e}}^{c_1} k_2^{b_1}B_2^{ed_1}
    -k_1^{d_1}{B_{1e}}^{a_1} k_2^{b_1}B_2^{c_1e}
    +{\textstyle\half} k_{1e}B_1^{a_1c_1} k_2^eB_2^{b_1d_1} \big).
\end{align}
Not all of these terms can be generated from new terms in the
effective action: those terms with `naked' $B$-fields, \ie\ those
without any momenta multiplying them such as the $B_1$ in the first
term, cannot be covariantised in a gauge-invariant manner. The
resolution is that the first four terms in \eqref{eq:BBLeftOverTerms}
and the two terms in \eqref{eq:BhLeftOverTerms} actually cancel when
summed over all permutations of the external states. The proof of this
uses an identity between four $\bar{t}_{10}$ tensors which is shown in
the appendix of \cite{Richards:2008jg}. Consider
\begin{align} \label{eq:t10identity}
  &(\bar{t}_{10}^{\,ABa_1b_1a_2b_2a_3b_3a_4b_4} +\bar{t}_{10}^{\,ABa_2b_2a_1b_1a_3b_3a_4b_4}
  +\bar{t}_{10}^{\,ABa_3b_3a_1b_1a_2b_2a_4b_4} +\bar{t}_{10}^{\,ABa_4b_4a_1b_1a_2b_2a_3b_3}) \notag\\
  &\quad\times t_8^{c_1d_1c_2d_2c_3d_3c_4d_4} B_{1AB} k_{2a_1}k_{2c_1}B_{2b_1d_1}
  k_{3a_2}k_{3c_2}h_{1b_2d_2} k_{4a_3}k_{4c_3}h_{2b_3d_3} k_{5a_4}k_{5c_4}h_{3b_4d_4},
\end{align}
which vanishes due to identity (A.4) in \cite{Richards:2008jg}. By
expanding each of the $\bar{t}_{10}$ tensors as a sum of four $t_8$
tensors as in (A.2) in \cite{Richards:2008jg}, it is easy to show that
this gives the first two terms in \eqref{eq:BBLeftOverTerms} and the
two terms in \eqref{eq:BhLeftOverTerms} summed over all permutations
of the gravitons. The final two `naked' $B$-terms in
\eqref{eq:BBLeftOverTerms} cancel with the $B_1\leftrightarrow B_2$
equivalent of \eqref{eq:BhLeftOverTerms}, which is shown by
considering the above identity but with $B_1$ and $B_2$ interchanged.

This leaves the final five terms in \eqref{eq:BBLeftOverTerms} and we
can now ask what term or terms in the effective action might give rise
to them. By considering
\begin{equation} \label{eq:H2R3term}
  \alpha'^3\int d^{10}x \sqrt{-g}\,
  t_8^{a_1b_1a_2b_2a_3b_3a_4b_4} t_8^{c_1d_1c_2d_2c_3d_3c_4d_4}
  H_{a_1c_1e} {H_{b_1d_1}}^e R_{a_2b_2c_2d_2} R_{a_3b_3c_3d_3}
  R_{a_4b_4c_4d_4}
\end{equation}
it is easy to see that, at lowest order in $h$, exactly these five
extra terms are generated and this then is our required new term to
account for the $B^2h^3$ one-loop amplitude.

This term agrees with the conjecture in \cite{Kehagias:1997cq} (see
their (2.11)), but disagrees with the covariant RNS calculation in
\cite{Peeters:2001ub}, where it is claimed that the $t_8t_8$ part of
the $H^2R^3$ term is as above but with the $H^2$ part replaced by
$H_{a_1b_1e} {H_{c_1d_1}}^e$. We believe the difference arises
because \cite{Peeters:2001ub}, as indeed acknowledged there, makes no
attempt to subtract field theory diagrams due to known quartic terms
in the effective action. It is also worth noting that our result
disagrees with the light-cone gauge GS calculation in
\cite{Frolov:2001xr}.

\section{The $B^4h$ Amplitude} \label{sec:B4h}
It is relatively straightforward to extend the whole of the previous
analysis to the $B^4h$ amplitude. In calculating the $B^2h^3$
amplitude in section \ref{sec:B2h3amp} (and in the five-graviton case
in \cite{Richards:2008jg}) at most two vertex operators give
$\partial X^a$ and $\bar\partial X^a$ terms, with the remaining
three operators giving $R^{ab}k^b\tilde{R}^{cd}k^d$ factors. These
three factors simply lead to $k_bk_dX_{ac}$ factors in the amplitude,
where $X$ is either $h$ or $B$, and so the difference between $B^2h^3$
and $B^4h$ is minimal. Let the four \NSNS\ two-forms be $B_1$, $B_2$,
$B_3$, $B_4$ with momenta $k_1$, $k_2$, $k_3$, $k_4$ respectively, and
let the graviton be labelled $h_1$ and have momentum $k_5$. Then the
$B^4h$ amplitude is given by \eqref{eq:amp1} but with $h^1h^2h^3$
replaced by $B^3B^4h^1$. Similarly, the lowest-order expansion is
given by the same replacement in \eqref{eq:amp2}.

\begin{figure}
\begin{center}
\includegraphics[scale=0.35]{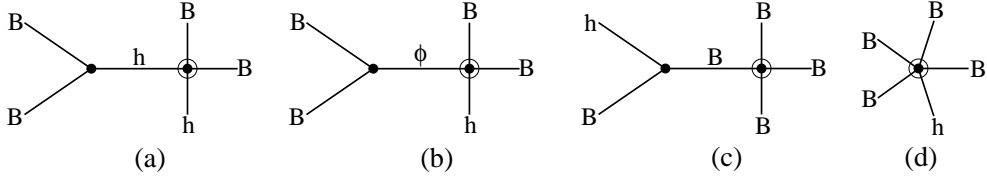}
\caption{Field theory diagrams contributing to the $B^4h$
  amplitude: (a)-(c) pole diagrams, and (d) a contact diagram.}
\label{fig:EffActDiags3}
\end{center}
\end{figure}

There is also little difference between the effective action diagrams
for $B^2h^3$ and $B^4h$. This follows since, as explained in
\cite{Richards:2008jg}, all diagrams, even the contact diagram,
contain exactly two particles which play a privileged r\^ole. The
remainder merely act as spectators, appearing as $R_{abcd}$ or
$\nabla_{[a}H_{b]cd}$ factors expanded to leading order. So there is
little point explicitly evaluating the new diagrams. However, for
completeness, the relevant diagrams are shown in figure
\ref{fig:EffActDiags3} where several new vertices are required: the
four-vertex from the lowest-order expansion of the $(\nabla H)^4$ term
in \eqref{eq:GrossSloanEA}, the equivalent vertex from the
$R\phi(\nabla H)^2$ term, and the five-vertex from the expansion of
$(\nabla H)^4$ to first order in $h$.

The poles in all channels match in an identical manner to that for the
$B^2h^3$ case and the `naked' $B$-fields cancel using the same
$\bar{t}_{10}$ identity. This leaves the same five $kBkB$ terms as for
$B^2h^3$, but now with two of the $k_ak_ch_{bd}$ factors replaced by
$k_ak_cB_{bd}$. So the new term required in the effective action is
given by
\begin{align} \label{eq:H2DH2Rterm}
  &\alpha'^3\int d^{10}x \sqrt{-g}\,
  t_8^{a_1b_1a_2b_2a_3b_3a_4b_4}t_8^{c_1d_1c_2d_2c_3d_3c_4d_4} \notag\\
  &\qquad\times H_{a_1c_1e} {H_{b_1d_1}}^e
  \nabla_{[a_2}H_{b_2]c_2d_2} \nabla_{[a_3}H_{b_3]c_3d_3}
  R_{a_4b_4c_4d_4}.
\end{align}

It is notable that, as in \cite{Gross:1986mw} where the quartic
effective action including $B$-fields was written by generalising
$R^4$ to $\bar{R}^4$, we can write both the $H^2R^3$ and $H^4R$ terms
as
\begin{equation} \label{eq:GrossSloanGen}
  \alpha'^3\int d^{10}x \sqrt{-g}\,
  t_8^{a_1b_1\cdots a_4b_4}t_8^{c_1d_1\cdots c_4d_4}
  H_{a_1c_1e} {H_{b_1d_1}}^e
  \bar{R}_{a_2b_2c_2d_2} \bar{R}_{a_3b_3c_3d_3}
  \bar{R}_{a_4b_4c_4d_4},
\end{equation}
where
\begin{equation} \label{eq:barR2}
  \bar{R}_{abcd} = R_{abcd} + {\textstyle\half}e^{-\phi/2}\nabla_{[a}H_{b]cd}.
\end{equation}
In fact, as in \eqref{eq:barR}, \cite{Gross:1986mw} also includes an
additional term in $\bar{R}_{abcd}$ which involves the dilaton. It is
interesting to conjecture that the same may also be true here. This
predicts various new terms, such as $H^2R^2(\nabla\nabla\phi)$ and
$H^4(\nabla\nabla\phi)^3$. To confirm such terms it would be necessary
to calculate five-particle amplitudes involving dilatons, such as the
$B^2h^2\phi$ amplitude.

\section{The $\epsilon_8\epsilon_8$ Terms} \label{sec:eeterms}
So far we have only considered $t_8$ tensors, where $t_8$ is formed
out of products of delta symbols \cite{Green:1987mn} and which
originate from a trace over four $R_0^{ab}$ tensors. This trace,
however, also contains an $\epsilon_8$ tensor,
\begin{align} \label{eq:trR4}
   \tr (R^{ab}_0 R^{cd}_0 R^{ef}_0 R^{gh}_0)
   &= \pm\half\epsilon_8^{abcdefgh}
   -\half\delta^{ac}\delta^{bd}\delta^{eg}\delta^{fh}
   + \cdots \notag\\
   &\equiv \pm\half\epsilon_8^{abcdefgh} + t_8^{abcdefgh},
\end{align}
with the $\pm$ sign depending on the $SO(8)$ chirality. Although
momentum conservation requires all terms involving $\epsilon_8$ to
vanish in massless \NSNS\ four-point amplitudes, this is no longer the
case for amplitudes involving five states. This leads to various terms
in the effective action such as $\epsilon_{10}\epsilon_{10}R^4$ and
$B\wedge t_8R^4$. Since the GS light-cone formalism requires $k^+=0$
for all external states, the $B\wedge t_8R^4$ term will not be
visible. However, terms with at least two contractions between the
epsilon tensors should be visible. The $\epsilon_{10}\epsilon_{10}R^4$
term was studied in \cite{Richards:2008jg}; here we study equivalent
terms involving $B$-fields, such as $\epsilon_{10}\epsilon_{10}(\nabla
H)^2R^2$ and $\epsilon_{10}\epsilon_{10}H^2R^3$.

All $t_8$ factors in the $B^2h^3$ amplitude originate from a trace
over four $R_0^{ab}$ and so $\epsilon_8$ terms can be included simply
by replacing $t_8$ by \eqref{eq:trR4}. Since the identities used to
reach \eqref{eq:amp1} continue to hold for the $\epsilon_8$ terms
\cite{Richards:2008jg}, the final amplitude expanded to lowest order
is given by \eqref{eq:amp2} but with every occurrence of $t_8$ in
\eqref{eq:A12} and \eqref{eq:B12} replaced by $\half\epsilon_8+t_8$.
As a consequence of the greater antisymmetry of $\epsilon_8$, all but
the final line of $A_{rs}$ vanish and so the $\epsilon_8$ parts of
the amplitude contain no massless poles.

In addition to the $t_8t_8$ terms studied above, the full amplitude
now contains $t_8\epsilon_8$ and $\epsilon_8\epsilon_8$ terms. The
$t_8\epsilon_8$ must vanish since they do not respect the parities of
either the IIA or IIB theory. For the five-graviton amplitude at
lowest order in $\alpha'$ this was demonstrated explicitly in
\cite{Richards:2008jg}; the equivalent statement for $B^2h^3$ can be
shown in an identical manner. So the full amplitude reduces to
$(t_8t_8\pm\quart\epsilon_8\epsilon_8)$ multiplied by the usual
kinematic factors and integrals, with $+/-$ for IIB/IIA
respectively. As discussed in \cite{Richards:2008jg}, this calculation
appears to give the wrong sign for the $\epsilon_8\epsilon_8$ terms,
as can be seen by comparing with the known
$\epsilon_{10}\epsilon_{10}R^4$ term in the effective action. We leave
this issue unresolved and instead conjecture that $+/-$ should refer
to IIA/IIB.

As with the $t_8t_8$ terms, before any $\epsilon_{10}\epsilon_{10}H^2R^3$
can be determined it is important to subtract diagrams due to quartic
terms in the effective action. For the pure Riemann term it is known
that the one-loop $t_8t_8R^4$ should be extended to
\begin{equation}
  \alpha'^3\int d^{10}x\sqrt{-g}\,e^{\frac{1}{2}\phi} 
  (t_8t_8\pm{\textstyle\frac{1}{8}}\epsilon_{10}\epsilon_{10})R^4,
\end{equation}
with $+/-$ for IIB/IIA respectively \cite{Kiritsis:1997em,
Antoniadis:1997eg}. It is natural to conjecture that the
$t_8t_8\bar{R}^4$ term in \eqref{eq:GrossSloanEA} is extended in a
similar way to 
$(t_8t_8\pm\frac{1}{8}\epsilon_{10}\epsilon_{10})\bar{R}^4$, giving
the new terms
\begin{equation} \label{eq:GrossSloanEpsilon}
  \pm\frac{1}{8}\alpha'^3\int d^{10}x\sqrt{-g}\,
  \epsilon_{10\,mn}^{\phantom{10\,mn}a_1b_1a_2b_2a_3b_3a_4b_4}
  \epsilon_{10}^{mnc_1d_1c_2d_2c_3d_3c_4d_4}
  \bar{R}_{a_1b_1c_1d_1}\cdots \bar{R}_{a_4b_4c_4d_4}.
\end{equation}
The following analysis will confirm this for a subset of these
terms, including $\epsilon_{10}\epsilon_{10}(\nabla H)^4$ and
$\epsilon_{10}\epsilon_{10}R^2(\nabla H)^2$.

The expansion of \eqref{eq:GrossSloanEpsilon} around Minkowski space
proceeds exactly as in section \ref{sec:EA}. The only new ingredient
is the expansion of the $\epsilon_{10}\epsilon_{10}$ factor. As
shown in \cite{Richards:2008jg}, with the understanding that
$\epsilon_{10}\epsilon_{10}$ is always multiplied by a tensor which is
symmetric under the interchange of pairs of adjacent indices, \ie\
under $(a_r,b_r) \leftrightarrow (a_s,b_s)$, then to first order in $h$,
\begin{align}
  {\epsilon_{10\,mn}^{}}^{a_1b_1\cdots a_4b_4}
  {\epsilon_{10}}^{mnc_1d_1\cdots c_4d_4} &\to -2
  \underline{\epsilon}_8^{a_1b_1\cdots a_4b_4}
  \underline{\epsilon}_8^{c_1d_1\cdots c_4d_4} \notag\\
  &\qquad + 8(
  {h_i}^{a_1}\underline{\epsilon}_8^{ib_1\cdots a_4b_4}
    \underline{\epsilon}_8^{c_1d_1\cdots c_4d_4} +
  {h_i}^{b_1}\underline{\epsilon}_8^{a_1i\cdots a_4b_4}
    \underline{\epsilon}_8^{c_1d_1\cdots c_4d_4} ).
\end{align}
Here $\epsilon_{10}\epsilon_{10}$ are `curved' epsilon tensors which
can be rewritten as a sum of products of the metric $g$, and
$\underline\epsilon_8\underline\epsilon_8$ are `flat' epsilon tensors
which can be rewritten using flat metrics $\eta$.

Due to the extra antisymmetry of the epsilon tensor, the four-vertex
from \eqref{eq:GrossSloanEpsilon} vanishes and, as a consequence, the
$\epsilon_{10}\epsilon_{10}$-equivalents of diagrams
\ref{fig:EffActDiags}(a)-(d) all vanish, leaving only diagram
\ref{fig:EffActDiags}(e). This tallies with the lack of massless poles
in the $\epsilon_8\epsilon_8$ part of the amplitude. However, since
the expansion of $\epsilon_{10}\epsilon_{10}$ is so similar to the
expansion of $t_8t_8$, it is prudent to ignore this fact. Then the
subtraction of the field theory diagrams from the amplitude proceeds
exactly as for the $t_8t_8$ terms.

After subtracting all diagrams and again using \eqref{eq:t10identity},
but with $t_8$ replaced by $\epsilon_8$, it is the same five
terms in the $\epsilon_8\epsilon_8$-equivalent of
\eqref{eq:BBLeftOverTerms} which remain, leading to two
conclusions. Firstly, since the field theory diagrams which must be
subtracted involve the $\epsilon_8\epsilon_8BBhh$ and
$\epsilon_8\epsilon_8hhh\phi$ vertices, we confirm the presence of
both the $\epsilon_{10}\epsilon_{10}R^2(\nabla H)^2$ and
$\epsilon_{10}\epsilon_{10}R^3\phi$ terms of
\eqref{eq:GrossSloanEpsilon}. Secondly, analogous to
\eqref{eq:H2R3term}, it is necessary to add the term
\begin{align} \label{eq:eeH2R3}
  &\pm\frac{1}{8}\alpha'^3\int d^{10}x \sqrt{-g}\,
  \epsilon_{10\,mn}^{\phantom{10\,mn}a_1b_1a_2b_2a_3b_3a_4b_4}
  \epsilon_{10}^{mnc_1d_1c_2d_2c_3d_3c_4d_4} \notag\\
  &\qquad\qquad\times
  H_{a_1c_1e} {H_{b_1d_1}}^e R_{a_2b_2c_2d_2} R_{a_3b_3c_3d_3}
  R_{a_4b_4c_4d_4},
\end{align}
where, as mentioned above, we switch the signs so that $+/-$ applies
to IIB/IIA respectively. Since this has the same structure as the
$t_8t_8H^2R^3$ term, the full $H^2R^3$ term can be written as the
combination
\begin{equation}
  \alpha'^3\int d^{10}x \sqrt{-g}\,(t_8t_8\pm{\textstyle\frac{1}{8}}
  \epsilon_{10}\epsilon_{10})H^2R^3,
\end{equation}
where $H^2R^3$ is shorthand for the above tensor contractions. This
mirrors the usual $t_8t_8R^4$ term, which is similarly generalised to
$(t_8t_8 \pm {\textstyle\frac{1}{8}}\epsilon_{10}\epsilon_{10})R^4$.

The $\epsilon_8$ terms for the $Bh^4$ amplitude work in an almost
identical manner. The field theory diagrams now contain two new
vertices, the $\epsilon_8\epsilon_8BBBB$ and
$\epsilon_8\epsilon_8hBB\phi$ vertices. From these we can confirm the
presence of both the $\epsilon_{10}\epsilon_{10}(\nabla H)^4$ and
$\epsilon_{10}\epsilon_{10}R(\nabla H)^2\phi$ terms of
\eqref{eq:GrossSloanEpsilon}. Further, the $t_8t_8H^2(\nabla H)^2R$
term needs to be supplemented by
\begin{align} \label{eq:eeH4R}
  &\pm\frac{1}{8}\alpha'^3\int d^{10}x \sqrt{-g}\,
  \epsilon_{10\,mn}^{\phantom{10\,mn}a_1b_1a_2b_2a_3b_3a_4b_4}
  \epsilon_{10}^{mnc_1d_1c_2d_2c_3d_3c_4d_4} \notag\\
  &\qquad\qquad\times H_{a_1c_1e} {H_{b_1d_1}}^e
  \nabla_{[a_2}H_{b_2]c_2d_2} \nabla_{[a_3}H_{b_3]c_3d_3}
  R_{a_4b_4c_4d_4},
\end{align}
and so, as with $H^2R^3$, the full $H^2(\nabla H)^2R$ term packages
together as $\alpha'^3\int d^{10}x\sqrt{-g}\,
(t_8t_8\pm{\textstyle\frac{1}{8}}\epsilon_{10}\epsilon_{10})
H^2(\nabla H)^2R$.

As mentioned in section \ref{sec:B4h}, it is again notable that the
full $H^2R^3$ and $H^2(\nabla H)^2R$ terms can both be written
succinctly as the single term
\begin{align}
  &\alpha'^3\int d^{10}x \sqrt{-g}\,
  (t_8^{a_1b_1\cdots a_4b_4}t_8^{c_1d_1\cdots c_4d_4}\pm
  {\textstyle\frac{1}{8}}\epsilon_{10\,mn}^{\phantom{10\,mn}a_1b_1\cdots
  a_4b_4}\epsilon_{10}^{mnc_1d_1\cdots c_4d_4}) \notag\\
  &\qquad\quad\times H_{a_1c_1e} {H_{b_1d_1}}^e
  \bar{R}_{a_2b_2c_2d_2} \bar{R}_{a_3b_3c_3d_3}
  \bar{R}_{a_4b_4c_4d_4},
\end{align}
with $\bar{R}$ given in \eqref{eq:barR2}.

\acknowledgments
I am very grateful to Michael Green, Hugh Osborn and Pierre Vanhove
for many useful discussions.

\bibliographystyle{utcaps}
\bibliography{references}

\providecommand{\href}[2]{#2}\begingroup\raggedright\begin{thebibliography}{10}

\bibitem{Richards:2008jg}
D.~M. Richards, ``{The One-Loop Five-Graviton Amplitude and the Effective
  Action},'' {\em JHEP} {\bf 10} (2008) 042,
\href{http://www.arXiv.org/abs/0807.2421}{{\tt 0807.2421}}.

\bibitem{Green:1997tv}
M.~B. Green and M.~Gutperle, ``Effects of D-instantons,'' {\em Nucl. Phys.}
  {\bf B498} (1997) 195--227,
\href{http://www.arXiv.org/abs/hep-th/9701093}{{\tt hep-th/9701093}}.

\bibitem{Peeters:2001ub}
K.~Peeters, P.~Vanhove, and A.~Westerberg, ``Chiral splitting and world-sheet
  gravitinos in higher- derivative string amplitudes,'' {\em Class. Quant.
  Grav.} {\bf 19} (2002) 2699--2716,
\href{http://www.arXiv.org/abs/hep-th/0112157}{{\tt hep-th/0112157}}.

\bibitem{Vafa:1995fj}
C.~Vafa and E.~Witten, ``A One loop test of string duality,'' {\em Nucl. Phys.}
  {\bf B447} (1995) 261--270,
\href{http://www.arXiv.org/abs/hep-th/9505053}{{\tt hep-th/9505053}}.

\bibitem{Gross:1986mw}
D.~J. Gross and J.~H. Sloan, ``The Quartic Effective Action for the Heterotic
  String,'' {\em Nucl. Phys.} {\bf B291} (1987)
41.

\bibitem{Green:1987mn}
M.~B. Green, J.~H. Schwarz, and E.~Witten, ``Superstring Theory. Vol. 2: Loop
  Amplitudes, Anomalies and Phenomenology,''. Cambridge, Uk: Univ. Pr. ( 1987)
  596 P. ( Cambridge Monographs On Mathematical Physics).

\bibitem{D'Hoker:1993ge}
E.~D'Hoker and D.~H. Phong, ``Momentum analyticity and finiteness of the one
  loop superstring amplitude,'' {\em Phys. Rev. Lett.} {\bf 70} (1993)
  3692--3695,
\href{http://www.arXiv.org/abs/hep-th/9302003}{{\tt hep-th/9302003}}.

\bibitem{D'Hoker:1993mr}
E.~D'Hoker and D.~H. Phong, ``Dispersion relations in string theory,'' {\em
  Theor. Math. Phys.} {\bf 98} (1994) 306--316,
\href{http://www.arXiv.org/abs/hep-th/9404128}{{\tt hep-th/9404128}}.

\bibitem{D'Hoker:1994yr}
E.~D'Hoker and D.~H. Phong, ``The Box graph in superstring theory,'' {\em Nucl.
  Phys.} {\bf B440} (1995) 24--94,
\href{http://www.arXiv.org/abs/hep-th/9410152}{{\tt hep-th/9410152}}.

\bibitem{Green:1999pv}
M.~B. Green and P.~Vanhove, ``The low energy expansion of the one-loop type II
  superstring amplitude,'' {\em Phys. Rev.} {\bf D61} (2000) 104011,
\href{http://www.arXiv.org/abs/hep-th/9910056}{{\tt hep-th/9910056}}.

\bibitem{Gross:1986iv}
D.~J. Gross and E.~Witten, ``Superstring Modifications of Einstein's
  Equations,'' {\em Nucl. Phys.} {\bf B277} (1986)
1.

\bibitem{Green:2008uj}
M.~B. Green, J.~G. Russo, and P.~Vanhove, ``{Low energy expansion of the
  four-particle genus-one amplitude in type II superstring theory},'' {\em
  JHEP} {\bf 02} (2008) 020,
\href{http://www.arXiv.org/abs/0801.0322}{{\tt 0801.0322}}.

\bibitem{Green:1997di}
M.~B. Green and P.~Vanhove, ``D-instantons, strings and M-theory,'' {\em Phys.
  Lett.} {\bf B408} (1997) 122--134,
\href{http://www.arXiv.org/abs/hep-th/9704145}{{\tt hep-th/9704145}}.

\bibitem{Green:1998by}
M.~B. Green and S.~Sethi, ``Supersymmetry constraints on type IIB
  supergravity,'' {\em Phys. Rev.} {\bf D59} (1999) 046006,
\href{http://www.arXiv.org/abs/hep-th/9808061}{{\tt hep-th/9808061}}.

\bibitem{Green:1997as}
M.~B. Green, M.~Gutperle, and P.~Vanhove, ``One loop in eleven dimensions,''
  {\em Phys. Lett.} {\bf B409} (1997) 177--184,
\href{http://www.arXiv.org/abs/hep-th/9706175}{{\tt hep-th/9706175}}.

\bibitem{Green:1987sp}
M.~B. Green, J.~H. Schwarz, and E.~Witten, ``Superstring Theory. Vol. 1:
  Introduction,''. Cambridge, Uk: Univ. Pr. ( 1987) 469 P. ( Cambridge
  Monographs On Mathematical Physics).

\bibitem{Sannan:1986tz}
S.~Sannan, ``Gravity as the Limit of the Type II Superstring Theory,'' {\em
  Phys. Rev.} {\bf D34} (1986)
1749.

\bibitem{Kehagias:1997cq}
A.~Kehagias and H.~Partouche, ``The exact quartic effective action for the type
  IIB superstring,'' {\em Phys. Lett.} {\bf B422} (1998) 109--116,
\href{http://www.arXiv.org/abs/hep-th/9710023}{{\tt hep-th/9710023}}.

\bibitem{Frolov:2001xr}
S.~Frolov, I.~R. Klebanov, and A.~A. Tseytlin, ``String corrections to the
  holographic RG flow of supersymmetric SU(N) x SU(N+M) gauge theory,'' {\em
  Nucl. Phys.} {\bf B620} (2002) 84--108,
\href{http://www.arXiv.org/abs/hep-th/0108106}{{\tt hep-th/0108106}}.

\bibitem{Kiritsis:1997em}
E.~Kiritsis and B.~Pioline, ``On R**4 threshold corrections in type IIB string
  theory and (p,q) string instantons,'' {\em Nucl. Phys.} {\bf B508} (1997)
  509--534,
\href{http://www.arXiv.org/abs/hep-th/9707018}{{\tt hep-th/9707018}}.

\bibitem{Antoniadis:1997eg}
I.~Antoniadis, S.~Ferrara, R.~Minasian, and K.~S. Narain, ``R**4 couplings in
  M- and type II theories on Calabi-Yau spaces,'' {\em Nucl. Phys.} {\bf B507}
  (1997) 571--588,
\href{http://www.arXiv.org/abs/hep-th/9707013}{{\tt hep-th/9707013}}.

\end{thebibliography}\endgroup

\end{document}